\documentclass[12pt]{article}
\usepackage{graphicx}
\usepackage{amssymb}
\usepackage{amsmath}
\usepackage{bm}

\newcommand\pubnumber{}
\newcommand\pubdate{\today}

\def\pavia{Istituto Nazionale Fisica Nucleare - Sezione di Pavia\\
via Bassi 6, I-27100 Pavia, ITALY}

\def\Title#1{\begin{center} {\Large #1 } \end{center}}
\def\Author#1{\begin{center}{ \sc #1} \end{center}}
\def\Address#1{\begin{center}{ \it #1} \end{center}}

\newcommand\pubblock{\rightline{\begin{tabular}{l} \pubnumber\\
         \pubdate  \end{tabular}}}
\newenvironment{Abstract}{\begin{quotation}  }{\end{quotation}}
\newenvironment{Presented}{\begin{quotation} \begin{center} 
             PRESENTED AT\end{center}\bigskip 
      \begin{center}\begin{large}}{\end{large}\end{center} \end{quotation}}

\def\beq{\begin{equation}}
\def\eeq#1{\label{#1}\end{equation}}
\def\eeqn{\end{equation}}

\def\beqa{\begin{eqnarray}}
\def\eeqa#1{\label{#1}\end{eqnarray}}
\def\eeqan{\end{eqnarray}}

\let\bar=\overbar

\def\etal{{\it et al.}}
\def\ie{{\it i.e.}}
\def\eg{{\it e.g.}}

\def\Dslash{\not{\hbox{\kern-4pt $D$}}}
\def\dslash{\not{\hbox{\kern-2pt $\del$}}}

\def\msb{{\bar{\ssstyle M \kern -1pt S}}}

\begin{document}
\begin{titlepage}
\pubblock

\vfill
\Title{Realistic estimate of valence transversity \\ from dihadron production}
\vfill
\Author{ Marco Radici}
\Address{\pavia}
\vfill
\begin{Abstract}
We have updated our extraction of the transversity parton distribution based on the analysis of pion-pair production in deep-inelastic scattering off transversely polarized targets in collinear factorization. The most recent COMPASS data for proton and deuteron targets, complemented by previous HERMES data on the proton, make it possible to perform a flavor separation of the valence components of the transversity distribution, using di-hadron fragmentation functions taken from the semi-inclusive production of two pion pairs in back-to-back jets in $e^+ e^-$ annihilation. The  $e^+ e^-$ data from BELLE have been reanalyzed to reach a more realistic estimate of the uncertainties on the chiral-odd interference fragmentation function. Our results represent the most accurate estimate of the uncertainties on the valence components of the transversity distribution currently available. 
\end{Abstract}
\vfill
\begin{Presented}
Twelfth Conference on the Intersections of Particle and Nuclear Physics (CIPANP 2015)\\
Vail, Colorado (USA),  May 19--24, 2015
\end{Presented}
\vfill
\end{titlepage}
\def\thefootnote{\fnsymbol{footnote}}
\setcounter{footnote}{0}

\section{Introduction}
\label{sec:intro}

Parton distribution functions (PDFs) describe combinations of number densities of quarks and gluons in a fast-moving hadron.  At leading twist, the quark structure of spin-half hadrons is described by three PDFs: the unpolarized distribution $f_1$, the longitudinal polarization (helicity) distribution $g_1$, and the transverse polarization (transversity) distribution $h_1$. From the phenomenological point of view, $h_1$ is the least known one because it is connected to QCD-suppressed processes where the parton helicity is flipped (\ie, it is a chiral-odd function). Therefore, it can be measured only in processes with two hadrons in the initial state (\eg, proton-proton collision) or one hadron in the initial state and at least one hadron in the final state (\eg, semi-inclusive DIS - SIDIS).  

By simultaneously fitting data on polarized single-hadron SIDIS and data on almost back-to-back emission of two hadrons in $e^+ e^-$ annihilations, the transversity distribution was extracted for the first time by the Torino 
group (for the latest release, see Ref.~\cite{Anselmino:2013}). The main difficulty of such analysis lies in the factorization framework used to interpret the data, since it involves Transverse Momentum Dependent PDFs (TMDs). QCD evolution of TMDs must be included to analyze SIDIS and $e^+ e^-$ data obtained at very different scales. But the computation is very difficult and only recently an attempt to give a (not complete) description of these effects was released~\cite{Kang:2014zza,Kang:2015msa}.

Alternatively, the transversity distribution can be extracted in the standard framework of collinear factorization using 
data on SIDIS with two hadrons detected in the final state. In fact, $h_1$ is multiplied with a specific chiral-odd Di-hadron Fragmentation Function (DiFF)~\cite{Collins:1994kq,Jaffe:1998hf,Radici:2001na}, which can be extracted from the corresponding $e^+ e^-$ annihilation process leading to two back-to-back pion pairs~\cite{e+e-:2003,noiBelle}. The collinear framework allows to keep under control the evolution equations of DiFFs~\cite{Ceccopieri:2007ip}. Using two-hadron SIDIS data on a proton target from 
HERMES~\cite{DiFFHERMES} and the BELLE data for the process $e^+ e^- \to (\pi^+ \pi^-) (\pi^+ \pi^-) X$~\cite{Vossen:2011fk}, the first extraction of the valence flavor combination $h_1^{u_v} - 1/4 \  h_1^{d_v}$ was performed point-by-point directly from the experimental bins~\cite{Bacchetta:2011ip}. By including also the COMPASS data for both proton and deuteron 
targets~\cite{Adolph:2012nw}, it became possible for the first time to separately parametrize the valence components of 
transversity~\cite{h1JHEP}. Recently, the analysis was updated~\cite{Radici:2015mwa} by selecting the more recent and more precise COMPASS data for a proton target~\cite{compass_2014}, as well as by refining the determination of the uncertainties on the extraction of DiFFs from $e^+ e^-$ BELLE data. 

In this contribution to the proceedings, we summarize the parametrization and the error analysis both for $h_1$ and for the DiFFs. 

\section{Theoretical framework}
\label{sec:theory}

We consider the process $\ell(k) + N(P) \to \ell(k') + H_1(P_1) + H_2(P_2) + X$, where $\ell$ denotes the beam lepton, $N$ the nucleon target with mass $M$ and polarization $S$, $H_1$ and $H_2$ the produced unpolarized hadrons with masses $M_1$ and $M_2$, respectively. We define the total $P_h = P_1 + P_2$ and relative $R = (P_1-P_2)/2$ momenta of the pair, with the invariant mass $P_h^2 = M_h^2 \ll Q^2=-q^2$ and $q = k - k'$ the momentum transferred. We define the azimuthal angles $\phi_R$ and $\phi_S$ as the angles of ${\bf R}_T$ and ${\bf S}_T$, respectively, around the virtual photon direction ${\bf q}$ (see Ref.~\cite{Gliske:2014wba} for a covariant definition). We also define the polar angle $\theta$ which is the angle between the direction of the back-to-back emission in the center-of-mass (cm) frame of the two hadrons, and the direction of $P_h$ in the photon-nucleon cm frame. Then, ${\bf R}_T = {\bf R} \sin\theta$ and 
$|{\bf R}|$ is a function of the invariant mass only~\cite{Bacchetta:2002ux}. Finally, we use the standard definition of the SIDIS invariants $x,\, y$; for a hadron pair, $z = z_1 + z_2$. To leading twist and for the collinear kinematics ${\bf P}_h \parallel {\bf q}$, the differential cross section for the two-hadron SIDIS off a transversely polarized nucleon target becomes~\cite{h1JHEP}
\begin{equation}
\frac{d\sigma}{dx \, dy\, dz\, d\phi_S\, d\phi_R\, d M_{h}^2\,d \cos{\theta}} =  \frac{\alpha^2}{x y\, Q^2}\, 
\Biggl\{ A(y) \, F_{UU}  + |{\bf S}_T|\, B(y) \, \sin(\phi_R+\phi_S)\,  F_{UT} \Biggr\} \; ,
\label{crossSIDIS}
\end{equation}
where $\alpha$ is the fine structure constant, $A(y) = 1-y+y^2/2$, $B(y) = 1-y$, and 
\begin{eqnarray} 
F_{UU} & = &x \sum_q e_q^2\, f_1^q(x; Q^2)\, D_1^q\bigl(z,\cos \theta, M_h; Q^2\bigr) \; , \nonumber \\
F_{UT} &=  &\frac{|{\bf R}| \sin \theta}{M_h}\, x\, 
\sum_q e_q^2\,  h_1^q(x; Q^2)\,H_1^{\sphericalangle\, q}\bigl(z,\cos \theta, M_h; Q^2\bigr) \; , 
\label{StructFunct}
\end{eqnarray}
with $e_q$ the fractional charge of a parton with flavor $q$. The $D_1^q$ is the DiFF describing the hadronization of an unpolarized parton with flavor $q$ into an unpolarized hadron pair. The $H_1^{\sphericalangle\, q}$ is its 
chiral-odd partner describing the same fragmentation but for a transversely polarized parton~\cite{Bianconi:1999cd}. DiFFs can be expanded in Legendre polynomials in $\cos \theta$~\cite{Bacchetta:2002ux}. After averaging over
$\cos \theta$, only the term corresponding to the unpolarized pair being created in a relative $\Delta L=0$ state survives in the $D_1$ expansion, while the interference in $|\Delta L| = 1$ survives for 
$H_1^{\sphericalangle}$~\cite{Bacchetta:2002ux}. Without ambiguity, the two terms will be identified with $D_1$ and $H_1^{\sphericalangle}$, respectively. 

Inserting the structure functions of Eq.~(\ref{StructFunct}) into the cross section~(\ref{crossSIDIS}), we get the single-spin asymmetry (SSA)~\cite{Radici:2001na,Bacchetta:2002ux,Bacchetta:2011ip,h1JHEP, Radici:2015mwa}
\begin{equation}
A_{{\rm SIDIS}}(x, z, M_h; Q) =  - \frac{B(y)}{A(y)} \,\frac{|\bm{R} |}{M_h} \, 
\frac{ \sum_q\, e_q^2\, h_1^q(x; Q^2)\, H_1^{\sphericalangle\, q}(z, M_h; Q^2)    } 
        { \sum_q\, e_q^2\, f_1^q(x; Q^2)\, D_{1}^q (z, M_h; Q^2) }\;  .
\label{SIDISssa}
\end{equation} 
For the specific case of production of $\pi^+ \pi^-$ pairs, isospin symmetry and charge conjugation suggest 
$D_1^q = D_1^{\bar{q}}$ and $H_1^{\sphericalangle\, q} = - H_1^{\sphericalangle\, \bar{q}}$, with $q=u,d,s$, 
with also $H_1^{\sphericalangle\, u} = - H_1^{\sphericalangle\, d}$~\cite{Bacchetta:2006un,Bacchetta:2011ip,h1JHEP, Radici:2015mwa}. So, the actual combination for the proton target is~\cite{Bacchetta:2011ip,h1JHEP, Radici:2015mwa}, 
\begin{eqnarray} 
x\, h_1^{p}(x; Q^2) &\equiv &x \, h_1^{u_v}(x; Q^2) - {\textstyle \frac{1}{4}}\, x h_1^{d_v}(x; Q^2) \nonumber \\
&= &-\frac{ A^p_{{\rm SIDIS}} (x; Q^2)  }{n_u^{\uparrow}(Q^2)}\,\frac{A(y)}{B(y)} \, \frac{9}{4} \sum_{q=u,d,s} \, e_q^2\, 
n_q (Q^2)\, x f_1^{q+\bar{q}}(x; Q^2) \; , 
\label{xh1p}
\end{eqnarray}
and, for the deuteron target~\cite{h1JHEP, Radici:2015mwa}, 
\begin{eqnarray} 
 x\, h_1^{D} (x; Q^2) &\equiv &x \, h_1^{u_v}(x; Q^2)+ x h_1^{d_v}(x; Q^2)   \nonumber \\
 &= &- \frac{A^D_{\text{SIDIS}}(x; Q^2)}{n_u^{\uparrow}(Q^2)} \,\frac{A(y)}{B(y)} \, 3 \sum_{q=u,d,s}\, 
 \big[ e_q^2\, n_q (Q^2) + e_{\tilde{q}}^2\, n_{\tilde{q}} (Q^2) \big] \, x f_1^{q+\bar{q}}(x; Q^2) \; , \nonumber \\
& &  \label{xh1D}
\end{eqnarray}
where $h_1^{q_v} \equiv h_1^q - h_1^{\bar{q}}$, $f_1^{q+\bar{q}} \equiv f_1^q + f_1^{\bar{q}}$, $\tilde{q}=d,u,s$ if $q=u,d,s$, respectively, and 
\begin{eqnarray} 
n_q(Q^2) &= &\int_{z_{{\rm min}}}^{z_{{\rm max}}} \int_{M_{h\, {\rm min}}}^{M_{h\, {\rm max}}} 
dz \, dM_h \, D_1^q (z, M_h; Q^2)  \nonumber  \\
n_q^\uparrow (Q^2) &= &\int_{z_{{\rm min}}}^{z_{{\rm max}}} \int_{M_{h\, {\rm min}}}^{M_{h\, {\rm max}}} 
dz \, dM_h \, \frac{|{\bf R}|}{M_h}\, H_1^{\sphericalangle\, q}(z,M_h; Q^2) \; .
\label{DiFFnq}
\end{eqnarray} 

\section{Extraction of Dihadron Fragmentation Functions}
\label{sec:DiFF}

The quantities in Eq.~(\ref{DiFFnq}) can be determined by extracting DiFFs from the $e^+ e^- \to (\pi^+ \pi^-) (\pi^+ \pi^-) X$ process. In fact, the leading-twist cross section in collinear factorization, namely by integrating upon all transverse momenta but ${\bf R}_T$ and 
${\bf \bar{R}}_T$, can be written as~\cite{noiBelle}
\begin{equation}
d\sigma = \frac{1}{4\pi^2}\, d\sigma^0 \, \bigg( 1+ \cos (\phi_R + \phi_{\bar{R}} ) \, A_{e+e-} \bigg) \; , 
\label{e+e-cross}
\end{equation}
where the azimuthal angles $\phi_R$ and $\phi_{\bar{R}}$ give the orientation of the planes containing the momenta of the pion pairs with respect to the lepton plane (see Fig.1 of Ref.~\cite{noiBelle} for more details), and we define the so-called Artru-Collins 
asymmetry~\cite{e+e-:2003}
\begin{equation}
A_{e+e-} \propto \frac{|{\bf R}_T|}{M_h} \, \frac{|{\bf \bar{R}}_T|}{\bar{M}_h} \, 
\frac{\sum_q e_q^2\, H_1^{\sphericalangle\, q}(z,M_h; Q^2)\, H_1^{\sphericalangle\, \bar{q}}(\bar{z},\bar{M}_h; Q^2)}{\sum_q e_q^2\, D_1^q(z,M_h; Q^2)\, D_1^{\bar{q}}(\bar{z},\bar{M}_h; Q^2)} \, .
\label{e+e-ssa}
\end{equation} 

\begin{figure}[htb]
\begin{center}
\includegraphics[height=4.5cm,width=9cm]{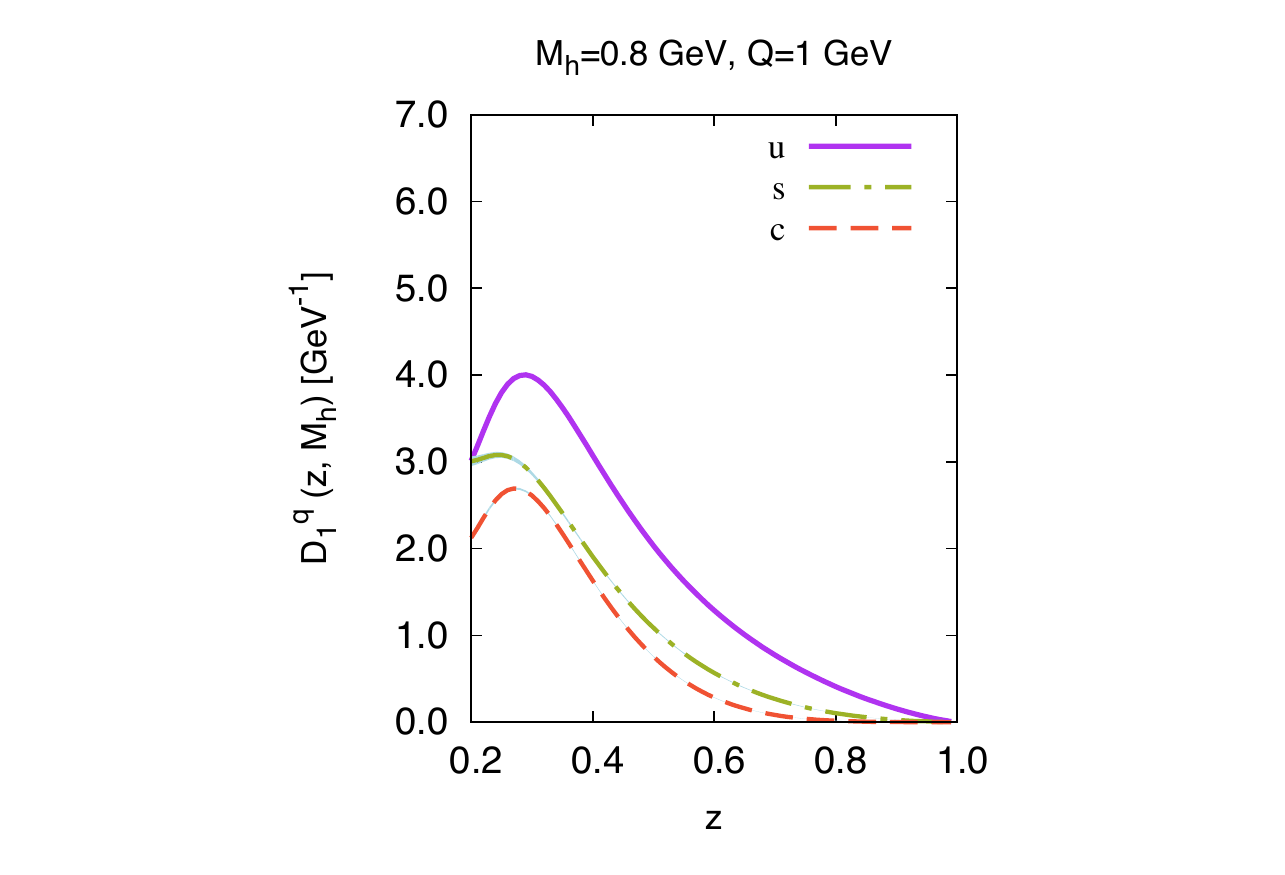}\hspace{-1.5cm}\includegraphics[width=6.5cm]{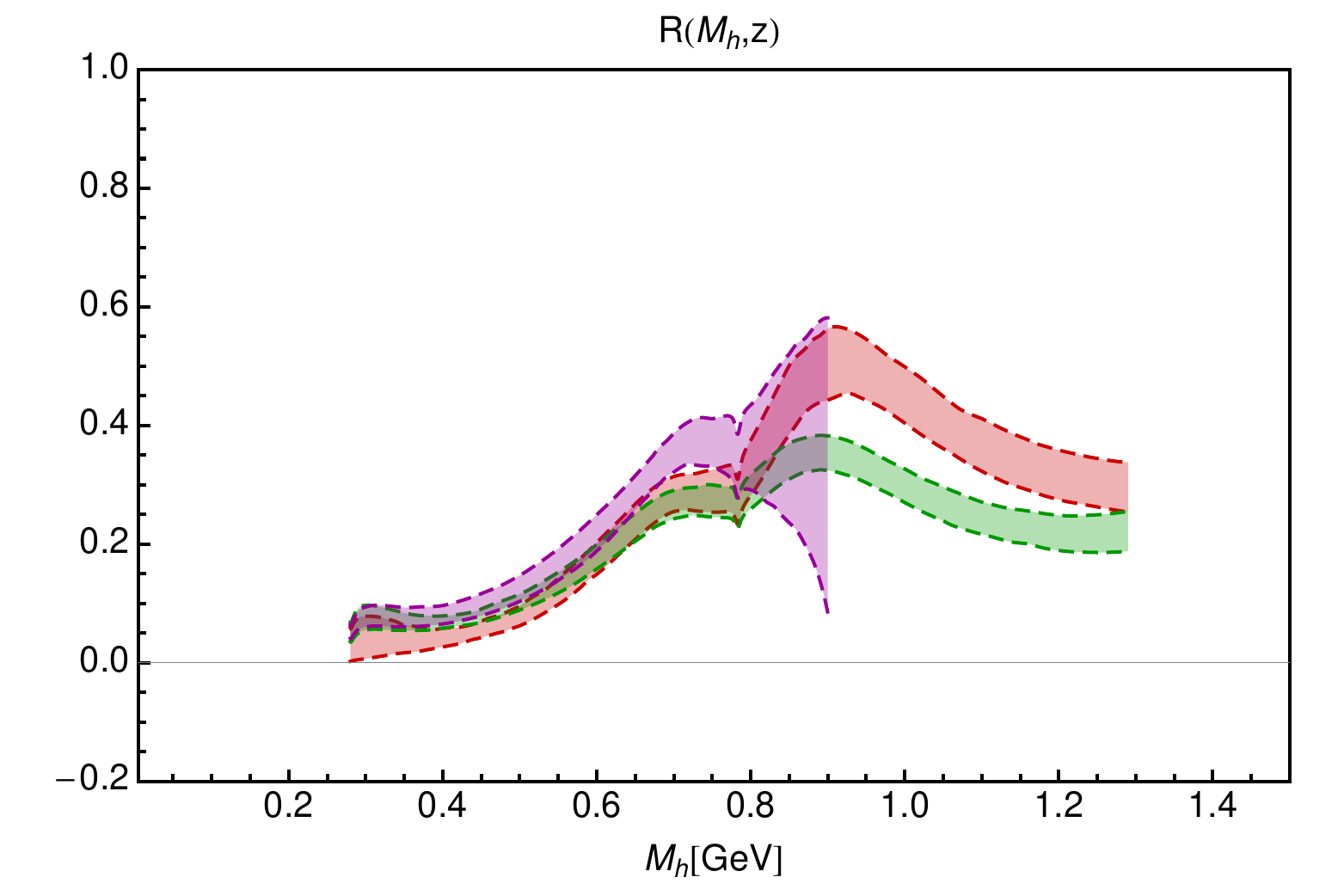}
\caption{Left panel: $D_1^q$ as a function of $z$ at $Q_0^2=1$ GeV$^2$ and $M_h=0.8$ GeV for $q=u (=d), s, c$. Right panel: 
$(|{\bf R}| / M_h) \, (H_1^{\sphericalangle\, u} / D_1^u)$ as a function of $M_h$ at $Q_0^2=1$ GeV$^2$ for three different $z=0.25,\, 0.45,\, 0.65$.}
\label{fig:DiFF}
\end{center}
\end{figure}

Since a measurement of the unpolarized $e^+ e^-$ cross section is still missing, the unpolarized DiFF $D_1$ was parametrized to reproduce a two-pion yield in about 32000 bins produced by the PYTHIA event generator tuned to the BELLE kinematics~\cite{noiBelle}. The fitting expression at the starting scale $Q_0^2=1$ GeV$^2$ was inspired by previous model 
calculations~\cite{Bacchetta:2006un,Radici:2001na,Bianconi:1999uc,SIDISe+e-2009} and it contains 3 resonant channels (pion pair produced by $\rho$, $\omega$, $K^0_S$ decays) and a continuum~\cite{noiBelle}. In the left panel of Fig.~\ref{fig:DiFF}, the $D_1^q$ for $q=u(=d),s,c$ flavors at $Q_0^2$ and $M_h=0.8$ GeV is shown as a function of $z$. 

Then, the chiral-odd $H_1^{\sphericalangle}$ was extracted from $A_{e+e-}$ by using the above mentioned isospin symmetry and charge conjugation of DiFFs and by integrating upon the hemisphere of the antiquark jet~\cite{noiBelle}. The experimental data are fitted starting from an expression for $H_1^{\sphericalangle\, u}$ at $Q_0^2=1$ GeV$^2$ with 9 parameters, and then evolving it to the BELLE scale. In Ref.~\cite{noiBelle}, the fit was performed with the traditional Hessian method reaching a final $\chi^2$/dof of 0.57. In Ref.~\cite{Radici:2015mwa}, the analysis was repeated using a different approach, which consists in perturbing the experimental points with a Gaussian noise to create $M$ replicas of them, and in separately fitting the $M$ replicas. The final outcome is a set of $M$ different fitting functions. The 68\% uncertainty band can be simply obtained by rejecting the largest and smallest 16\% of values for each experimental bin. The value of $M$ is determined by accurately reproducing the mean and standard deviation of the original data points; in this case, $M=100$. In the right panel of Fig.~\ref{fig:DiFF}, the ratio $(|{\bf R}|/M_h) \, (H_1^{\sphericalangle\, u}/D_1^u)$ at $Q_0^2=1$ GeV$^2$ is reported as a function of $M_h$ for three different $z=0.25,\, 0.45,\, 0.65$.  

\section{Extraction of Transversity}
\label{sec:h1}

The combination of Eqs.~(\ref{xh1p}) and (\ref{xh1D}) makes it possible to separately parametrize each valence flavor of the transversity distribution. The main theoretical constraint on transversity is the Soffer's inequality~\cite{Soffer:1995ww}. We impose this condition by multiplying the functional form by the corresponding Soffer bound $\mbox{\small SB}^q(x; Q^2)$ at the starting scale $Q_0^2=1$ GeV$^2$ (the explicit expression for SB$^q$ can be found in the Appendix of Ref.~\cite{h1JHEP}). Our analysis is carried out at LO in $\alpha_S$, whose normalization at the $Z$ boson mass ($\alpha_S (M_Z^2)$) is varied in order to account for the theoretical uncertainty in the determination of the $\Lambda_{\rm QCD}$ parameter. The functional form reads
\begin{equation} 
x\, h_1^{q_v}(x; Q_0^2)=
\tanh \Bigl[ x^{1/2} \, \bigl( A_q+B_q\, x+ C_q\, x^2+D_q\, x^3\bigr)\Bigr]\, x \, 
\Bigl[ \mbox{\small SB}^q(x; Q_0^2)+ \mbox{\small SB}^{\bar q}(x; Q_0^2)\Bigr] \, .
\label{funct_form}
\end{equation} 
The hyperbolic tangent is such that the Soffer bound is always fulfilled. The low-$x$ behavior is determined by the $x^{1/2}$ term, which is imposed by hand to grant the integrability of Eq.~(\ref{funct_form}). Present fixed-target data do not allow to constrain it. The functional form can contain up to 3 nodes. Here, we show the results employing only the parameters $A, B, C$, the so-called {\it flexible} scenario. 

The error analysis was carried out in a way similar to the reanalysis of DiFFs described in the previous section, namely using the more general replica method. 

\begin{figure}[htb]
\begin{center}
\includegraphics[width=6cm]{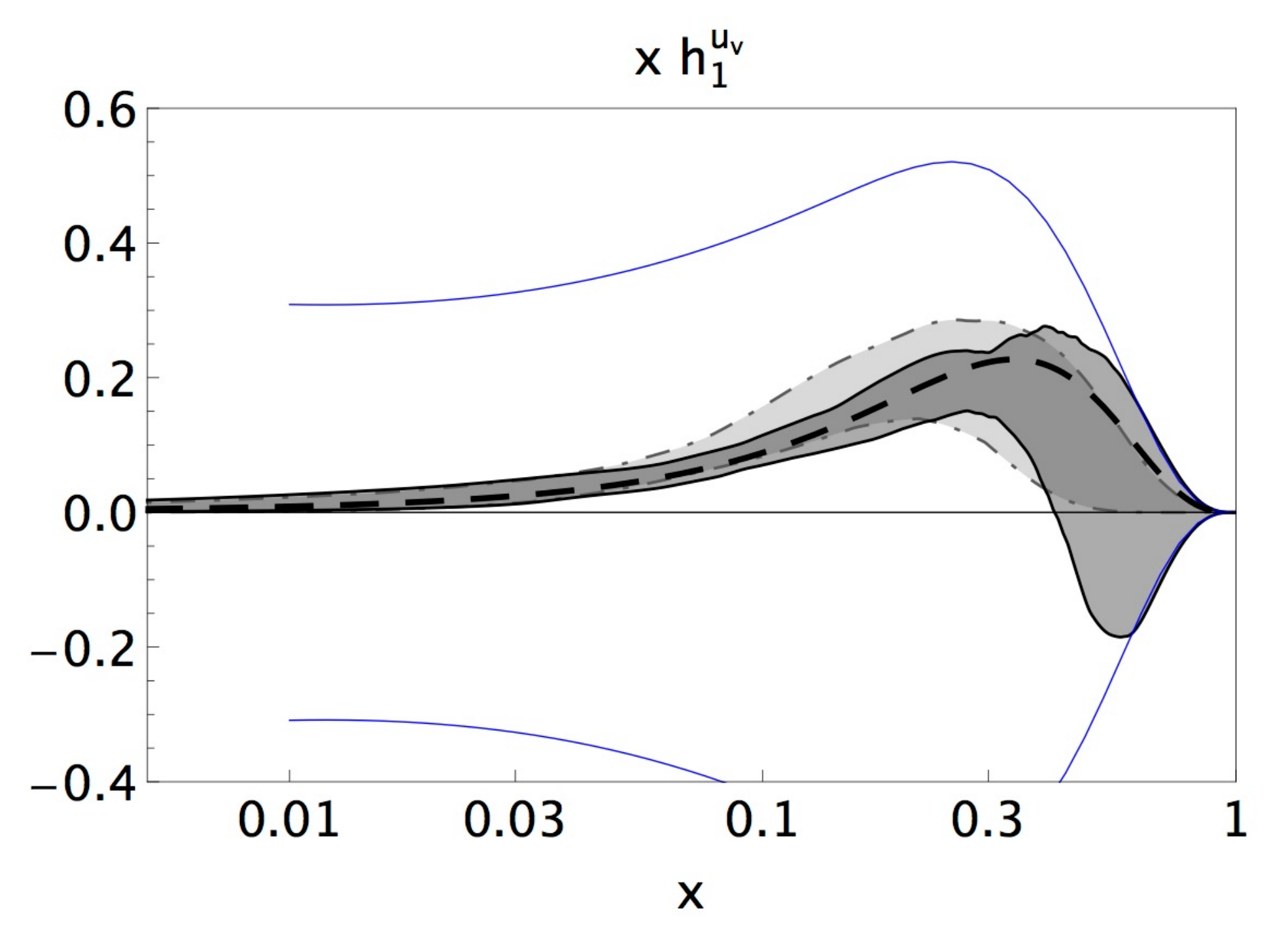}\hspace{0.5cm}\includegraphics[width=6cm]{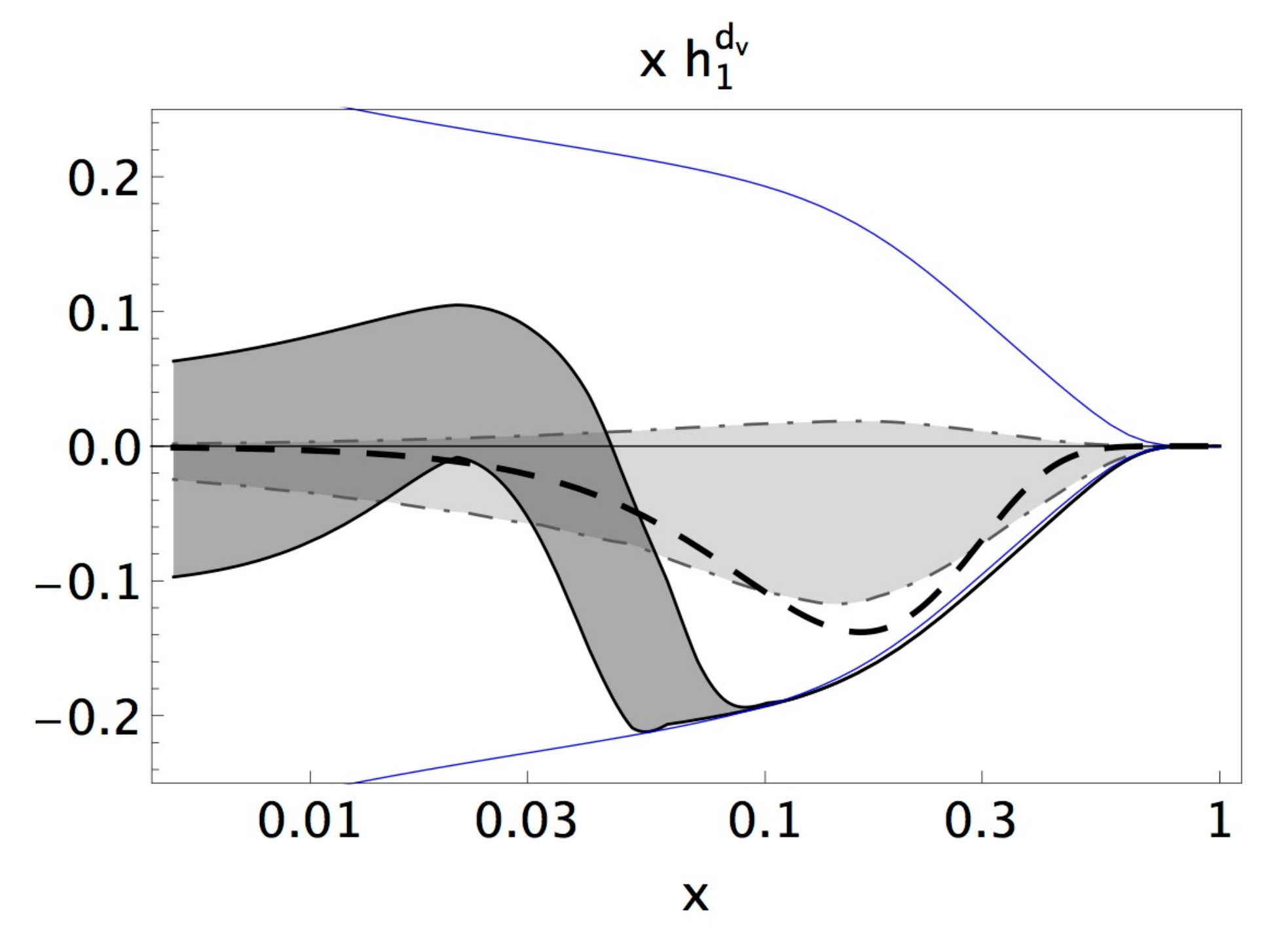}
\end{center}
\caption{Left panel: the transversity $x h_1^{u_v} (x)$ at $Q^2=2.4$ GeV$^2$ in the {\it flexible} scenario. The darker band with solid borders in the foreground is the result of Ref.~\cite{Radici:2015mwa} with $\alpha_S (M_Z^2) = 0.125$~\cite{Gluck:1998xa}. The lighter band with dot-dashed borders in the background is the most recent extraction from the Collins effect~\cite{Anselmino:2013}. The central thick dashed line is the result of Ref.~\cite{Kang:2014zza}. The dark thick solid lines indicate the Soffer bound. Right panel: same notations for the $d_v$ component.}
\label{xh1}
\end{figure}

In Fig.~\ref{xh1}, the left panel displays the transversity $x h_1^{u_v} (x)$ at the scale $Q^2 = 2.4$ GeV$^2$ for the {\it flexible} scenario, while $x h_1^{d_v} (x)$ is in the right one. For each panel, the darker band with solid borders is  the 
68\% of all the $M=100$ replicas obtained in Ref.~\cite{Radici:2015mwa} with $\alpha_S (M_Z^2) = 0.125$~\cite{Gluck:1998xa}. The lighter band with dot-dashed borders is the most recent transversity extraction from the Collins effect~\cite{Anselmino:2013}. The central thick dashed line is the result of Ref.~\cite{Kang:2014zza}, where evolution equations have been computed in the TMD framework. This analysis has been recently updated~\cite{Kang:2015msa} including also a calculation of the error band which turns out to mostly overlap with the lighter band from Ref.~\cite{Anselmino:2013}. Finally, the dark thick solid lines indicate the Soffer bound. 

There is consistency among the various extractions, at least for $0.0065 \leq x \leq 0.29$ where there are data. However, the error analysis based on the replica method gives a more realistic description of the uncertainty on transversity, specifically for large $x$ outside the data range. As it is clear in the left panel of Fig.~\ref{xh1}, for $x \geq 0.3$ the replicas tend to fill all the phase space available within the Soffer bound. In order to reduce this uncertainty, it is important that new data will be collected in this region with the forthcoming upgrade of Jefferson Lab to the 12 GeV beam. In the right panel, for $x \geq 0.1$ our results tend to saturate the lower limit of the Soffer bound because they are driven by the COMPASS deuteron data, in particular by the bins number 7 and 8. This happens for all the explored scenarios, indicating that it is not an artifact of the chosen functional form. No such trend is evident in the parametrization corresponding to the single-hadron measurement from the Collins effect. It is also interesting to remark that the dashed line from Ref.~\cite{Kang:2014zza}, although in general agreement with the other extraction based on the Collins effect, also tends to saturate the Soffer bound at $x > 0.2$. 

The first Mellin moment of the transversity gives the tensor charge of the nucleon. We can give a reliable estimate for the tensor charge by truncating the integral to the $x$--interval where there are data. For the {\it flexible} scenario and with $\alpha_S (M_Z^2) = 0.125$, we find at $Q^2=10$ GeV$^2$ that $\delta u_v = 0.25 \pm 0.05$ and $\delta d_v = -0.25 \pm 0.12$~\cite{Radici:2015mwa}. These results are stable with respect to other scenarios and choices of $\alpha_S (M_Z^2)$, and, within errors, are compatible with the extraction from the single-hadron Collins effect in the TMD framework~\cite{Kang:2015msa}. We also extended the integration to $0\leq x \leq 1$ by extrapolating the fitting function. The result is influenced by the choice of the low-$x$ tail, which is unconstrained by data.  Anyway, within the (large) error bars there is consistency between our calculations and, \eg, the results from the Collins effect of Ref.~\cite{Anselmino:2013}.

\section{Universality of transversity}
\label{sec:pp}

The agreement displayed in Fig.~\ref{xh1} among the various extractions of transversity is a first important cross-check, but the actual verification of transversity being a universal parton distribution implies to make predictions for different processes involving transversity at different energies. In the $p p^\uparrow \rightarrow (h_1\ h_2) X$ process, a proton with momentum $P_A$ collides on a transversely polarized proton with momentum $P_B$ and spin vector $S_B$, producing a pair of unpolarized hadrons $h_1, \, h_2,$ inside the same jet. The transverse component of the total pair momentum ${\bf P}_h$ with respect to the beam ${\bf P}_A$ is indicated with ${\bf P}_{h\perp}$ and serves as the hard scale of the process. If the kinematics is collinear, namely if the transverse component ${\bf P}_{hT}$ of ${\bf P}_h$ around the jet axis is integrated over, the differential cross section at leading order in $1/|{\bf P}_{h\perp}|$ is~\cite{Bacchetta:2004it}
\begin{equation}
\frac{d\sigma}{d\eta\, d|{\bf P}_{h\perp}|\, dM_h^2\,d\phi_R} =  
 d\sigma^0 \  \left( 1 + \sin (\phi_{S_B} - \phi^{}_R) \, A_{pp} \right)  
\label{ppcross}
\end{equation}
with $\phi_{S_B} = \pi / 2$, where 
\begin{equation}
d\sigma^0=  2 \, |{\bf P}_{h\perp}| \, \sum_{a,b,c,d}\int \frac{d x_a\, dx_b }{4 \pi^2 z_h} \, f_1^a (x_a) \, f_1^b(x_b) \, \frac{d\hat{\sigma}_{ab \to cd}}{d\hat{t}} \, D_1^c (z_h, M_h^2) 
\label{ppcross0}
\end{equation}
and 
\begin{equation}
A_{pp}= [d\sigma^0]^{-1}\, 2 \, |{\bf P}_{h\perp}|\, \frac{|{\bf R}|}{M_h}\,|{\bf S}_{BT}|\, \sum_{a,b,c,d}\, \int \frac{dx_a \, dx_b}{16 \pi z_h} 
\, f_1^a(x_a) \, h_1^b(x_b) \, \frac{d\Delta \hat{\sigma}_{ab^\uparrow \to c^\uparrow d}}{d\hat{t}} H_1^{\sphericalangle c}(z_h, M_h^2) \, .
\label{App}
\end{equation}
Again, the angle $\phi_R$ describes the azimuthal orientation around ${\bf P}_h$ of the plane containing the hadron pair momenta, but now with respect to the $( {\bf P}_h, \, {\bf P}_A )$ reaction plane. Moreover, the pseudorapidity $\eta$ is defined by~\cite{Bacchetta:2004it}
\begin{equation}
z_h = \frac{|{\bf P}_{h\perp}|}{\sqrt{s}}\, \frac{x_a e^{-\eta} + x_b e^{\eta}}{x_a x_b} \, , 
\label{rapidity}
\end{equation}
where $\sqrt{s}$ is the cm energy of the collision, and $\hat{t} = t \  x_a / z_h$, with $t = (P_A - P_B)^2$. All possible combinations of partons $a + b \rightarrow c + d,$ must be included, they are described by the unpolarized and polarized cross sections $d\hat{\sigma}$ and $d\Delta\hat{\sigma}$, respectively (see Ref.~\cite{Bacchetta:2004it} for the complete list). 

\begin{figure}[htb]
\begin{center}
\includegraphics[width=7cm]{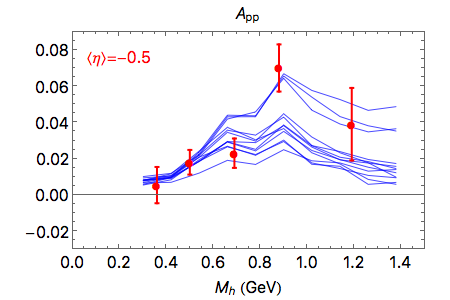}
\end{center}
\caption{The spin asymmetry $A_{pp}$ of Eq.~(\ref{App}) as function of $M_h$ for $\langle \eta \rangle = -0.5$ and integrated in $|{\bf P}_{h\perp}|$. Data from STAR~\cite{Adamczyk:2015hri} at $\sqrt{s} = 200$ GeV. Lines are preliminary results for 12 replicas of $h_1$ and $H_1^{\sphericalangle}$ from Ref.~\cite{Radici:2015mwa}.}
\label{f::App}
\end{figure}

The asymmetry $A_{pp}$ of Eq.~(\ref{App}) has been measured by the STAR Collaboration for the process $p p^\uparrow \rightarrow (\pi^+ \pi^-) X$ at the cm energy of $\sqrt{s} = 200$ GeV~\cite{Adamczyk:2015hri}. In Fig.~\ref{f::App}, it is shown as a function of $M_h$ after integrating over $|{\bf P}_{h\perp}|$ and at the average pseudorapidity $\langle \eta \rangle = -0.5$, which corresponds to large $x$ in the valence region. Hence, these data add a complementary and very useful information to what we already know on transversity from the SIDIS analysis. The solid lines represent a preliminary calculation of $A_{pp}$ using a sample of 12 replicas for $h_1^q$ and $H_1^{\sphericalangle\, q}$ obtained from the SIDIS and $e^+ e^-$ analyses described in the previous sections, respectively. The preliminary nature of the calculation prevents from drawing any conclusion, but the agreement displayed in Fig.~\ref{f::App} is definitely encouraging. 

%

%

\end{document}